# Minimally Invasive Randomization for Collecting Unbiased Preferences from Clickthrough Logs

**Filip Radlinski** and **Thorsten Joachims**
Department of Computer Science
Cornell University, Ithaca, NY
{filip,tj}@cs.cornell.edu

## Abstract

Clickthrough data is a particularly inexpensive and plentiful resource to obtain implicit relevance feedback for improving and personalizing search engines. However, it is well known that the probability of a user clicking on a result is strongly biased toward documents presented higher in the result set irrespective of relevance. We introduce a simple method to modify the presentation of search results that provably gives relevance judgments that are unaffected by presentation bias under reasonable assumptions. We validate this property of the training data in interactive real world experiments. Finally, we show that using these unbiased relevance judgments learning methods can be guaranteed to converge to an ideal ranking given sufficient data.

## Introduction

The problem of learning to rank using relevance judgments has recently received significant attention within the machine learning community (for example (Cohen, Shapire, & Singer 1999; Chu & Keerthi 2005; Yu, Yu, & Tresp 2005; Radlinski & Joachims 2005)). Learning ranking functions is especially appealing for information retrieval on specialized collections or for specific communities of users where manual tuning of ranking algorithms is impractical. However, obtaining useful training data is difficult.

One option is to employ experts to provide judgments as to the relevance of particular documents to particular queries. For example, this method is often used in the information retrieval community. While it provides clean training data, it is usually prohibitively expensive and time-consuming due to the need to employ human experts.

An alternative approach is to derive relevance judgments from the behavior of normal users. This can yield virtually unlimited amounts of data at almost no cost, and the data reflects the judgments of the users of the search engine rather than a select group of experts who may have a different concept of relevance. Clicks on search results (commonly called clickthrough data) are the most easily observed user behavior in web interfaces to retrieval systems. While some researchers have considered collecting data from other behavioral cues such as view time or scrolling behavior (Kelly & Teevan 2003; Fox *et al.* 2005;

White, Ruthven, & Jose 2005), we focus on clickthrough data due to its simplicity and easy availability.

The concern with clickthrough data is that it is noisy and biased. It is noisy in the sense that different users may have different concepts of relevance even given the same query (Teevan, Dumais, & Horvitz 2005). It is biased in the sense that the user's choice whether or not to click on a result depends on a combination of at least the document relevance and its position in the search results (Joachims *et al.* 2005). This makes the process of separating the bias from the actual document relevance difficult, both from a practical and a theoretical perspective.

In this paper, we observe that through experiment design it is possible to modify the presentation of search results with the purpose of collecting cleaner training data from regular users, while having minimal effect on the quality of the results presented. In particular, we propose the FairPairs algorithm. In its simplest form, FairPairs flips adjacent pairs of results in the ranking presented to the user according to a randomized scheme. We show that it allows clickthrough data to provide relevance judgments that are unaffected by presentation bias. Given two reasonable assumptions, we prove that preferences collected with FairPairs are not affected by presentation bias. We verify the validity of the assumptions empirically using a real world search engine. We also show that FairPairs agrees with manual relevance judgments in a setting where we know the relative relevance of documents. Additionally, we prove that learning with data collected by FairPairs will eventually converge to an ideal ranking, if one exists, thus justifying our method as providing suitable training data. Using the FairPairs algorithm, it is also possible to measure the confidence that a particular pair of results is in the correct order. We believe that the idea of minimally invasive randomization is also relevant to other applications outside of web search, where training data is collected from non-experts and suffers from biases and noise. We start by reviewing related work in the next section, then we motivate and present the FairPairs algorithm. After showing the theoretical properties of FairPairs, we finish by providing empirical evidence supporting our approach.

## Training Data Alternatives

Training data for learning to rank can be obtained in at least three ways. The most straightforward is through the use of human experts. In the Text REtrieval Conference (TREC), ranked retrieval systems are evaluated by manually assessing documents in the top $N$ results returned by any retrieval system taking part in an evaluation experiment. These results are considered by human judges and rated either relevant or not relevant (Voorhees 2004). This process provides high quality training data, but is expensive and hence not practical in most situations.

A less expensive method to get significant training data even in small domains is to interpret a user clicking on web search results as a sign of relevance. For example, Kemp and Ramamohanarao (2002) used this approach in a university search engine, where documents clicked on in the search results were assumed to be relevant. They used document transformation to allow the search system to learn. However, Joachims et al. (2005) demonstrated that such clickthrough data is strongly biased by the position at which results are presented. In particular, they showed that if a search result is moved higher in the result set, this immediately increases the probability users will click on it, even if the result is not relevant to the query. Our experiments confirmed that those findings also hold outside of a laboratory setting.

A third approach that avoids this bias problem is to interpret user clicks as relative feedback about pairs of results (Joachims 2002; Radlinski & Joachims 2005). This work proposed to interpret clicks on results as a relative statement that a clicked result is likely more relevant than others skipped over earlier. This compensates for presentation bias by considering the order results are observed by users. However, the effect is that preferences always oppose the presented ordering. In particular, such relevance judgments are all satisfied if the ranking is reversed, making the preferences difficult to use as training data.

From a theoretical standpoint, many researchers have considered the question of stability and convergence of a learned ranking given some data collection method as the amount of training data grows (Freund *et al.* 1998; Herbrich, Graepel, & Obermayer 2000; Cohen, Shapire, & Singer 1999; Crammer & Singer 2001; Chu & Keerthi 2005). Most of this research has been for problems in ordinal regression, which considers the problem of learning to map items to a partial order and does not apply directly to more general ranking problems. In particular, it does not consider how user behavior biases the training data.

Finally, we view our approach as experiment design (see for example Hinkelmann & Kempthorne 1994). In traditional experiment design, the researcher asks the question of what to measure to ensure conclusive and unbiased results. In a web-based search engine, we view the presentation of results to users as part of an interactive process that can also be designed to provide unbiased data for machine learning purposes. For this reason, we consider the data collection phase as part of the learning process.

## Presentation Bias

We now introduce the concept of presentation bias. In normal web search, users pay significantly more attention to results ranked highly than those ranked lower. For example, we implemented a search engine (presented in the experi-mental results section) for the arXiv e-print archive, a large collection of academic articles. We observed that users click on the fifth ranked result for only about 5% of queries, and click on lower ranked results even less often. However, if we take the fiftieth result and place it first or second in the ranking, users click on it more than 5% of the time. Does this indicate that we have a poor ranking function where the fiftieth result tends to be more relevant than the fifth? Rather, this demonstrates the concept of presentation bias in clickthrough data, even for search engines where users tend to be academic researchers.

| Relative relevance | "Normal" $d_1$ | $d_2$ | "Swapped" $d_1$ | $d_2$ |
|---|---|---|---|---|
| $rel(d_1) > rel(d_2)$ | 20/36 | 2/36 | 16/28 | 2/28 |
| $rel(d_1) < rel(d_2)$ | 7/20 | 4/20 | 12/36 | 9/36 |

Table 1: Results from a user study using the Google search engine. In the "normal" condition, straight Google results were presented, while the top two results were swapped in the "swapped" condition. The counts show how often each result was clicked on when the Google's top result was more or less relevant.

Similarly, Joachims et al. (2005) performed a controlled user study where volunteer subjects were asked to search for specific information using Google. The results viewed by the subjects were afterward assessed by expert judges for relevance. Table 1 shows a small selection of the results. The subjects saw one of two experimental conditions. In the "normal" condition, the results were presented as ranked by Google. When the result presented at the top ($d_1$) was judged by a human expert to be more relevant than the result presented next ($d_2$), users clicked on the top result 20 out of 36 times and on the second result twice, as could be expected. However, in the "swapped" condition the top two results from Google were reversed before being presented to users. Even when the second-ranked result was more relevant, users still clicked predominantly on the top ranked result. This again shows that presentation strongly influences user behavior.

**Definition 1.** *Presentation Bias is bias present in user's decisions of whether or not to click on a search engine result based on the position of the result rather than on its relevance to the user's information need.*

Presentation bias may occur for a number of reasons, such as users trusting the search engine to always present the most relevant result first. The question we address is how to tease out information about the relevance of the search results from clickthrough logs despite such effects.

## Bias-Free Feedback

In this section, we review the notion of relative relevance preferences and then present the FairPairs algorithm. Training data for learning to rank can be represented either as *absolute* or as *relative* relevance statements. The former involve data of the form $relevance(doc_i \mid query) = r_i$ where $r_i$ is an absolute measure of relevance. This approach requires an absolute relevance scale in the training data, for

1. Let $R = (d_1, d_2, \ldots, d_n)$ be the results for some query.
2. Randomly choose $k \in \{0, 1\}$ with uniform probability.
3. If $k = 0$
   - For $i \in \{1, 3, 5, \ldots\}$
     - Swap $d_i$ and $d_{i+1}$ in $R$ with 50% probability.
4. Otherwise ($k = 1$)
   - For $i \in \{2, 4, 6, \ldots\}$
     - Swap $d_i$ and $d_{i+1}$ in $R$ with 50% probability.
5. Present $R$ to the user, recording clicks on results.
6. Every time the lower result in a pair that was considered for flipping is clicked, record this as a preference for that result over the one above it.

Table 2: The FairPairs algorithm.

example specifying $r_i \in [0, 1]$. In this situation, it is particularly difficult to obtain well calibrated partial relevance judgments: For example, in ranking movies from 1 to 5 stars, different judges interpret a rating of 3 starts differently. Instead we consider relative statements, with training data in the form of preferences such as $relevance(doc_i \mid query) > relevance(doc_j \mid query)$. The aim is to obtain judgments where the probability some $doc_i$ is judged more relevant than some $doc_j$ is independent of the ranks at which they are presented.

We now present FairPairs by example, then provide the formal algorithm. The key idea is to randomize part of the presentation to eliminate the effect of presentation bias while making only minimal changes to the ranking. Consider some query that returns the documents $(d_1, d_2, d_3, d_4, d_5, \ldots)$. We perturb the result set so that we can elicit relevance judgments unaffected by presentation bias. We first randomly pick $k \in \{0, 1\}$. If $k = 0$, we consider the result set as pairs $((d_1, d_2), (d_3, d_4), (d_5, d_6), \ldots)$. Each pair of results is now independently flipped with 50% probability. For example, the final ranking might end up as $(d_1, d_2, d_4, d_3, d_5, \ldots)$ with only $d_3$ and $d_4$ flipped. Alternatively, we could end up flipping all the pairs: Each time FairPairs is executed, a different reordering may occur. Then we take the result set generated in this way and present it to the user. In expectation half the results will be presented at their original rank, and all results will be presented within one rank of their original position. Similarly, if $k = 1$, we do the same thing except consider the result set as pairs $(d_1, (d_2, d_3), (d_4, d_5), \ldots)$. The FairPairs algorithm is formally presented in Table 2.

To interpret the clickthrough results of FairPairs, consider the results for some query $q$ that returns $(d_1, d_2, \ldots, d_n)$. Let $d_j \triangleleft d_i$ denote that $d_j$ is presented just above $d_i$ (i.e., the user sees $d_j$ first if they read from the top) and that $k$ is such that $d_i$ and $d_j$ are in the same pair (e.g., when $k = 0$, $d_3$ and $d_4$ are in the same pair, but $d_2$ and $d_3$ are not). Let $n_{ij}$ count of how often this occurs. Also, let $c_{ij}$ denote the number of times a user clicks on $d_i$ when $d_j \triangleleft d_i$ (i.e., $d_i$ is the bottom result in a pair). By perturbing the results with FairPairs, we have designed the experiment such that we can interpret $c_{ij}$ as the number of votes for $relevance(d_i) > relevance(d_j)$, and $c_{ji}$ as the number of votes for $relevance(d_j) > relevance(d_i)$. The votes are counted only if the results are presented in equivalent ways, providing an unbiased set of preferences because both sets of votes are affected by presentation bias in the same way. We formalize this property and prove its correctness in the next section. Note that if a user clicks multiple times on some set of results, they are making multiple votes.

Although in this paper we focus on preferences generated from user clicks on the bottom result of a pair, in fact most of the properties discussed also appear to hold for preferences generated from clicks on the top result of a pair. The reason we chose to focus on clicks on bottom results is that Granka (2004; 2004) showed in eye tracking studies that users typically read search engine results top to bottom and are less likely to look at the result immediately below one they click on than they are to look at one immediately above it.

## Theoretical Analysis

In this section we will show that given any presentation bias that satisfies two simple assumptions, FairPairs is guaranteed to give preference data that is unaffected by presentation bias.

We start by presenting our assumptions. Let $r_i(q)$ be the relevance of document $d_i$ to a query $q$ (we will usually omit $q$ for brevity). The probability of a particular document being clicked by a user depends on its position in the search results, its relevance to the query, as well as potentially on every other document presented to the user. Assume the user selects $d_{bot}$ from the list $(\mathbf{d}_\uparrow, d_{top}, d_{bot}, \mathbf{d}_\downarrow)$, where $\mathbf{d}_\uparrow$ are the documents preceding (ranked above) $d_{top}$ and $\mathbf{d}_\downarrow$ are those after (ranked below) $d_{bot}$. In particular, $d_{top}$ is the document just before $d_{bot}$. Let

$$P(d_{bot}|\mathbf{d}_\uparrow, (d_{top}, d_{bot}), \mathbf{d}_\downarrow)$$

be the probability that $d_{bot}$ is clicked by the user given the list of choices.

**Assumption 1 (Document Identity).** *The probability of a user clicking depends only on the relevance of the documents presented, not their particular identity. Formally, we can write this as*

$$P(d_{bot}|\mathbf{d}_\uparrow, (d_{top}, d_{bot}), \mathbf{d}_\downarrow) = P(d_{bot}|\mathbf{r}_\uparrow, (r_{top}, r_{bot}), \mathbf{r}_\downarrow)$$

This assumption essentially states that the user is looking for any sufficiently relevant document. It also requires that users do not choose to skip over documents they recognize and know to be relevant. We come back to this later.

We now define two scores. First, the item relevance score measures how much more likely users are to click on more relevant results.

**Definition 2 (Item Relevance Score).** *If we take a ranking of documents and replace some document $d_1$ with a less relevant one $d_2$ while leaving all others unchanged, the difference between the probability that $d_1$ being selected and that of $d_2$ being selected is the item relevance score. Formally, if $d_1$ and $d_2$ have relevance $r_1$ and $r_2$ with $r_1 > r_2$, then*

$$\delta_{12}^{rel} = P(d_1|\mathbf{r}_\uparrow, (r_{top}, r_1), \mathbf{r}_\downarrow) - P(d_2|\mathbf{r}_\uparrow, (r_{top}, r_2), \mathbf{r}_\downarrow)$$

Analogously, consider the effect of replacing the document before the one that the user selects.

**Definition 3 (Ignored Relevance Score).** *If we take a ranking of documents and replace some document with a more relevant one while leaving all others unchanged, the difference between the probability of the user selecting the next document (after the one replaced) and the same probability without the change is the ignored relevance score. Formally, if $d_1$ and $d_2$ have relevance $r_1$ and $r_2$ where $r_1 > r_2$, then*

$$\delta_{12}^{ign} = P(d_{bot}|\mathbf{r}_\uparrow, (r_1, r_{bot}), \mathbf{r}_\downarrow) - P(d_{bot}|\mathbf{r}_\uparrow, (r_2, r_{bot}), \mathbf{r}_\downarrow)$$

This score measures how replacing the previous document changes the probability of a user clicking on a result. If $\delta_{12}^{ign}$ is negative, it means that replacing the previous document with a more relevant one reduces the probability of users clicking on the document under consideration. While we may expect this to be the case, it is possible that $\delta_{12}^{ign}$ is positive: if a user sees a very irrelevant document, they may be more likely to give up and not even evaluate the next result presented. On the other hand, if a user sees a somewhat relevant document, they may be more inclined to consider further results. We will measure this later.

Our second assumption relates to the relative magnitude of these two scores. Note that it would be trivially satisfied if the first is positive and the second negative.

**Assumption 2 (Relevance Score Assumption).** *The item relevance score is larger than the ignored relevance score.*

$$\delta_{ij}^{rel} > \delta_{ij}^{ign}$$

We will evaluate the validity of our assumptions in the experimental results section. Also, note that they are satisfied by many common item selection models (e.g., users selecting results where they judge their probability of success above some threshold (Miller & Remington 2004)).

We will now prove that the data collected in this way is unaffected by presentation bias. Theorem 1 tells us that if the documents before and after a pair being considered vary independently of how the pair is ordered, observing that the expectation of the users' probability of selecting $d_i$ when $d_j \triangleleft d_i$ is higher than the expectation of the users' probability of selecting $d_j$ when $d_i \triangleleft d_j$ is both necessary and sufficient to deduce that $r_i > r_j$.

**Theorem 1.** *Let $d_i$ and $d_j$ be two documents with relevance $r_i$ and $r_j$. If assumptions 1 and 2 are satisfied and $P(\mathbf{r}_\uparrow, \mathbf{r}_\downarrow|d_i \triangleleft d_j) = P(\mathbf{r}_\uparrow, \mathbf{r}_\downarrow|d_j \triangleleft d_i)$ then $r_i > r_j \Leftrightarrow P_{ij} > P_{ji}$, where $P_{ij} = E_{\mathbf{r}_\uparrow, \mathbf{r}_\downarrow}[P(d_i|\mathbf{r}_\uparrow, (r_j, r_i), \mathbf{r}_\downarrow)]$.*

*Proof.* We start by rewriting the expectations of the probabilities and simplifying:

$$P_{ij} = \sum_{\mathbf{r}_\uparrow, \mathbf{r}_\downarrow} P(d_i|\mathbf{r}_\uparrow, (r_j, r_i), \mathbf{r}_\downarrow) P(\mathbf{r}_\uparrow, \mathbf{r}_\downarrow|d_i \triangleleft d_j)$$

$$P_{ji} = \sum_{\mathbf{r}_\uparrow, \mathbf{r}_\downarrow} P(d_j|\mathbf{r}_\uparrow, (r_i, r_j), \mathbf{r}_\downarrow) P(\mathbf{r}_\uparrow, \mathbf{r}_\downarrow|d_i \triangleleft d_j)$$

$$P_{ij} - P_{ji} = \sum_{\mathbf{r}_\uparrow, \mathbf{r}_\downarrow} [P(d_i|\mathbf{r}_{ij}) - P(d_j|\mathbf{r}_{ji})] P(\mathbf{r}_\uparrow, \mathbf{r}_\downarrow|d_i \triangleleft d_j)$$

where $\mathbf{r}_{ij} = (\mathbf{r}_\uparrow, (r_j, r_i), \mathbf{r}_\downarrow)$. Say $P_{ij} - P_{ji}$ is positive. The sum can be positive if and only if the first term is positive for at least one $\mathbf{r}_\uparrow$ and $\mathbf{r}_\downarrow$.

Applying Assumption 2, we see $r_i > r_j$ is equivalent to

$$P(d_i|\mathbf{r}_\uparrow, (r_j, r_i), \mathbf{r}_\downarrow) > P(d_j|\mathbf{r}_\uparrow, (r_i, r_j), \mathbf{r}_\downarrow)$$

This equivalence means that if the first term in the summation is positive, then $r_i > r_j$. Hence $P(d_i|\mathbf{r}_\uparrow, (r_j, r_i), \mathbf{r}_\downarrow) > P(d_j|\mathbf{r}_\uparrow, (r_i, r_j), \mathbf{r}_\downarrow)$ for all $\mathbf{r}_\uparrow$ and $\mathbf{r}_\downarrow$ so the first term must always be positive. The same applies if the difference is negative. Hence the difference in expectations on the number of clicks always has the same sign as the difference in document relevance. □

The theorem tells us that we can collect relevance judgments about many pairs in the result set at the same time by independently randomly reordering pairs, as is the case with FairPairs.

## Practical Considerations

We now discuss how the data collected using FairPairs is affected by variations between search engines and in user behavior, which gives rise to practical issues that should be kept in mind. The first effect to note is that prior to deciding whether to click, users only observe the abstracts presented by the search engine. A less relevant document presented with a misleadingly appealing abstract may generate more clicks that one that is more relevant but has a less appealing abstract. While addressed by Assumption 2, in practice this requires the search engine to generate snippets in an unbiased way, where the quality of a snippet does not vary differently for different documents with the types of queries entered by users. An alternative would be to consider user dwell time on results in addition to clickthrough. Different search engines may also have users who are more or less prepared to click on results ranked highly irrespective of the abstract. However this is not a concern as both documents within a pair are always ranked highly equally often and hence benefit from this trust equally. Other presentation effects, such as a bias against users clicking on documents that are not visible unless the user scrolls also do not introduce bias into the training data, as shown in the results.

Another issue to consider is that of user familiarity with results: documents that are known to be relevant by the user but not clicked on may collect fewer votes than would be expected. However, it has been established that users often revisit web pages, suggesting that this is not a concern. Nevertheless, on specific collections for specific user groups this may be a limitation. Similarly, if the relevance of documents evolves over time, data collected may become out of date, although this is true for any data collection method. Finally, one well known user behavior that FairPairs does not exploit is that of query reformulation (Radlinski & Joachims 2005). FairPairs does not allow preferences to be generated in a fair way between documents returned by sequential queries, although extending it in this way is an interesting challenge.

## Learning Convergence

In this section, we consider the convergence properties of a learning algorithm that minimizes the error rate trained on data collected with FairPairs. For simplicity, assume that no two documents have the same relevance to a query.

**Theorem 2.** *Let $n_{ij}$ be the number of times the user saw $d_j \triangleleft d_i$ and $c_{ij}$ be the number of times a user clicked on $d_i$ in this situation. Let $\epsilon = \frac{1}{2}\min_{i,j}|P_{ij} - P_{ji}|$.*

*Assume we have collected enough data using FairPairs such that $\forall d_i, d_j, |1 - n_{ji}/n_{ij}| < \epsilon$ and $|p_{ij} - P_{ij}| < \frac{1}{2}\epsilon$, where $p_{ij} = c_{ij}/n_{ij}$. Moreover, assume there exists a ranking function $f^*$ that ranks the documents perfectly in terms of decreasing relevance. Then, a learning algorithm that minimizes error rate will return $f^*$.*

*Proof.* Assume for the purpose of a contradiction that the learning algorithm learns a ranking function $f \neq f^*$ that has a lower error rate on the training data. Since the rankings differ, there must be at least one pair of documents $d_i$, $d_j$ where the rankings disagree. Assign $d_i$ to be such that $r_i > r_j$, which is equivalent to $d_i$ being returned higher than $d_j$ by the ranking function $f^*$.

The number of violated constraints involving $d_i$ and $d_j$ for the ranking function $f^*$ is

$$\begin{aligned}err^*_{ij} &= n_{ij}p_{ij}\mathbb{1}[rank^*(d_i) > rank^*(d_j)] \\ &+ n_{ji}p_{ji}\mathbb{1}[rank^*(d_j) > rank^*(d_i)] \\ &= n_{ji}p_{ji},\end{aligned}$$

where $\mathbb{1}$ is the indicator function and $rank^*(d_i)$ is the rank at which $d_i$ is returned by $f^*$. The number of violated constraints involving $d_i$ and $d_j$ for the ranking function $f$ is:

$$\begin{aligned}err^f_{ij} &= n_{ij}p_{ij}\mathbb{1}[rank^f(d_i) > rank^f(d_j)] \\ &+ n_{ji}p_{ji}\mathbb{1}[rank^f(d_j) > rank^f(d_i)] \\ &= n_{ij}p_{ij},\end{aligned}$$

since we know that $rank^f$ disagrees with $rank^*$ on the order of $d_i$ and $d_j$. By assumption, we know that

$$(1+\epsilon)n_{ij} > n_{ji} > (1-\epsilon)n_{ij}$$

Next, by Theorem 1, $P_{ij} > P_{ji}$ since $r_i > r_j$. The definition of $\epsilon$ implies $P_{ij} - P_{ji} \geq 2\epsilon$. Since we know that $|p_{ij} - P_{ij}| < \frac{1}{2}\epsilon$ and similarly for $p_{ji}$, we get $p_{ij} - p_{ji} > \epsilon$.

Pulling this all together,

$$\begin{aligned}err^f_{ij} - err^*_{ij} &= n_{ij}p_{ij} - n_{ji}p_{ji} \\ &> n_{ij}[(p_{ij} - p_{ji}) - \epsilon p_{ji}] \\ &> n_{ij}[\epsilon - \epsilon p_{ji}] \geq 0\end{aligned}$$

Since the difference in the number of violated constraints is zero for pairs where $f$ and $f^*$ agree, and positive for all others, the error rate of the learned function $f$ must be higher than that of $f^*$, meaning we have a contradiction. □

Note that the data collected by Joachims (2002) does not have this property of eventual convergence because it tends to learn to reverse any presented ranking.

To ensure eventual convergence, we now need to ensure that sufficient data is collected about every pair of documents so that $\forall d_i, d_j, |1-n_{ji}/n_{ij}| < \epsilon$ and $|p_{ij}-P_{ij}| < \frac{1}{2}\epsilon$. The first condition is eventually satisfied since each flip is

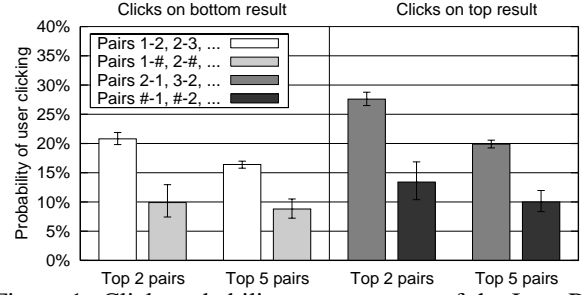

Figure 1: Click probability measurement of the Item Relevance Score.

performed with 50% probability. The second becomes satisfied after a pair is observed sufficiently often, because the probability of observing a click approaches its expectation by the law of large numbers. One strategy to obtain sufficient data would be for the search engine to occasionally insert random documents into the result set, and to assume users have a non-zero probability of viewing results at any rank. While this strategy would work (proof omitted due to space constraints), there are probably much more efficient exploration strategies, providing an interesting area for future research. Additionally, although the theorem is stated in terms of observing the relative relevance of all pairs of documents, if we assume relevance is transitive then the number of observations necessary may be substantially reduced.

## Experimental Results

In this section, we evaluate the validity of the assumptions presented and measure the click probabilities of data collected with FairPairs. For these experiments we used Osmot, a full text search engine, on the arXiv e-print collection consisting of about 350,000 academic articles. Osmot was modified to perform FairPairs on the search results before presenting them to real users who were unaware of the experiment. We measured the probability of users clicking on both the top and bottom results of each pair. We recorded user clicks for about three months, and counted how often each pair of ranks was presented and how often each result was clicked on. During our experiments, we observed 44,399 queries coming from 13,304 distinct IP addresses. We recorded 48,976 clicks on results, often with many clicks for the same query.

Because we do not have manual relevance judgments, we hypothesize that on average the fiftieth ranked result returned by the search system is less relevant to the query than the top few results. To check if we could confirm this, after FairPairs was performed on the results of a query, Osmot randomly swapped result fifty and one of the top eight results whenever there were more than fifty results for a query. This modified result set was then displayed to users.

### Item Relevance Score

The left side of Figure 1 shows how often users clicked on the bottom result of a pair. Four types of pairs were observed. Pairs of the form 1-2, 2-3 involve two adjacent results from the original ranking function in their original order (1-2 indicates it was the original first and second results, in the original order). Pairs of the form 2-1, 3-2 involve two originally adjacent results in reverse order. Due to the fiftieth result being randomly inserted, we also have pairs of the form 1-# (indicating the first result followed by the 50th result) and #-1 (indicating the same pair reversed). We summed up the counts for all pairs in these four groups, either for the top two pairs presented (e.g., for 1-2 and 2-3) or for the top five pairs (e.g., for 1-2 through 5-6) counting over all queries where a user clicked on at least one result. In the figure, we see that if the lower result in a pair is result 50 (postulated to be less relevant than those in the top six), the probability of the user clicking on that lower result is smaller than if the lower result was from the original top six. The error bars indicate the 95% binomial confidence intervals, showing the differences to be statistically significant. This shows that the Item Relevance Score is positive and gives an idea of its average magnitude at different ranks for this dataset. In fact, the right side of Figure 1 shows that keeping the lower result fixed, a similar score could be defined for the change in click probability on the top result as it is more or less relevant.

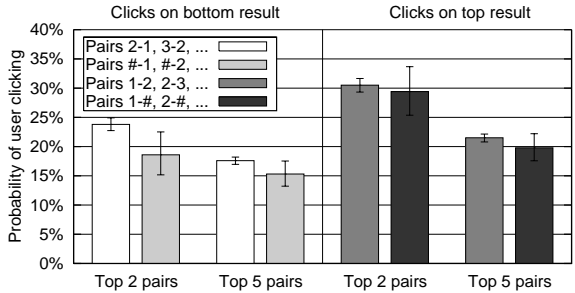

Figure 2: Click probability measurements of the Ignored Relevance Score.

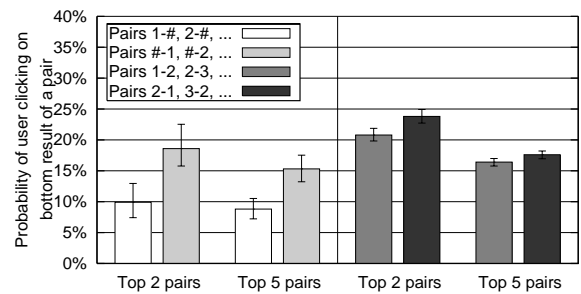

Figure 3: Evaluation of the relative relevance of search results returned by the arXiv search engine.

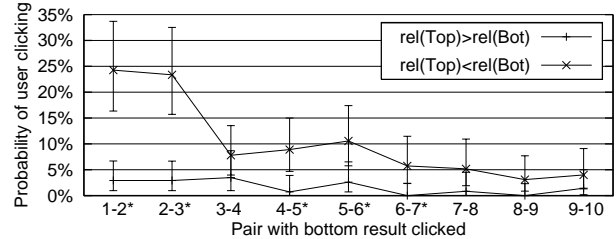

Figure 4: Probability of user clicking only on the bottom result of a pair as a function of the pair. The two curves are for when the document immediately above the document clicked was judged strictly more relevant or strictly less relevant by expert human judges. * indicates the difference is statistically significant with 95% confidence using a Fisher Exact test.

### Ignored Relevance Score

The ignored relevance score measures the change in click probability as the result *before* the one clicked on varies. We see in the left side of Figure 2 that if the result before the one selected is more relevant, the next document is slightly more likely to be clicked on. We attribute this to result 50 tending to be much less relevant, making users more likely to stop considering results once they encounter it. This means that in fact $\delta_{ij}^{ign}$ tends to be positive. Nevertheless, the magnitude of the decrease in click probability is much smaller than that seen in Figure 1, thus the Relevance Score Assumption holds. Additionally, we observe that this score quickly decreases for lower results unlike the item relevance score. Also, this is consistent with the right hand side of Figure 1 because there we saw the the probability of the user clicking on the top result whereas here we are evaluating the probability of the user clicking on the bottom result. Together, these figures show that placing a less relevant document as the top result in a pair makes both results in the pair less likely to be clicked on.

### FairPairs Preference Test

Next, to confirm the correctness of data generated with Fair-Pairs directly, consider the difference between the bottom click probabilities when two results are swapped. The left side of Figure 3 shows that reversing a top-five result and the 50th result within a pair behaves as the theory tells us it should. We see that when the fiftieth result is at the bottom of a pair, it is significantly less likely to be clicked on than when an original top-five result is at the bottom of the pair. On the right side of the figure, we see the click probability on the bottom result for pairs of the form 1-2 and for pairs of the form 2-1. In fact, summing the counts for the top 2 pairs (1-2 and 2-3), the difference in click probability is statistically significant. This shows that on average the top three results returned by the search engine are ranked in the correct order.

We also evaluated our approach in a situation where we have the true relative relevance of documents as assessed by human judges. Using the results of the study discussed earlier (Joachims *et al.* 2005), we computed the probability of a participant in the user study clicking on the bottom result of a pair of results when the top result was judged strictly more relevant or strictly less relevant by expert human judges. Figure 4 shows that although FairPairs was not performed on the results in the study, the data supports the FairPairs premise that the probability of a user clicking on a document $d_i$ at rank $i$ is higher if $rel(d_{i-1}) < rel(d_i)$ than if $rel(d_{i-1}) > rel(d_i)$.

Figure 5 shows the equivalent curve for the arXiv search engine, in effect providing a more detailed view of Figure 3. We again considered all queries that generated at least one click and exploited symmetries in our experiment design to obtain the maximal amount of data for this figure. It shows that if the fiftieth ranked document is displayed in a pair with a top-eight document, the FairPairs data collected is in agreement with our hypothesis that the fiftieth ranked document is less relevant than any from the top eight. In particular, the first five differences in click probabilities are statistically significant. For lower ranks the curves appear to proceed in a similar manner. This includes result pairs below the sixth, which are usually are not visible without users scrolling.

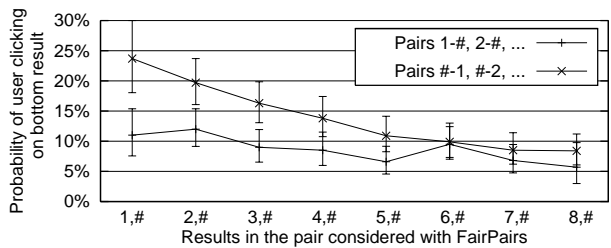

Figure 5: Probability of user clicking on the bottom result of a pair as a function of the pair for all queries generating at least one user click. The two curves are the cases where the document that was originally ranked fiftieth is the top or the bottom document in the pair. The error bars indicate 95% confidence intervals.

## Conclusions

In this paper we introduced FairPairs, a method to modify the presentation of search engine results with the purpose of collecting more reliable relevance feedback from normal user behavior. We showed that under reasonable assumptions the data gathered is provably unaffected by presentation bias. We also showed that given sufficient clickthrough data, training data generated with FairPairs will allow a learning algorithm to converge to the ideal ranking. We performed real world experiments that evaluated the assumptions and conclusions in practice. Given bias-free training data generated in this way, it is possible to use existing methods for learning to rank without additional modifications to compensate for presentation bias being necessary.

## Acknowledgments

We thank Thomas Finley, Eric Breck, Alexandru Niculescu-Mizil and the anonymous reviewers for helpful comments, as well as Simeon Warner and Paul Ginsparg for support with experiments performed using the arXiv e-print archive. This work was funded under NSF CAREER Award 0237381 and a research gift from Google, Inc.

## References

Chu, W., and Keerthi, S. S. 2005. New approaches to support vector ordinal regression. In *Proceedings of International Conference on Machine Learning (ICML)*.

Cohen, W. W.; Shapire, R. E.; and Singer, Y. 1999. Learning to order things. *Journal of Artificial Intelligence Research* 10:243–270.

Crammer, K., and Singer, Y. 2001. Pranking with ranking. In *Proceedings of the conference on Neural Information Processing Systems (NIPS)*.

Fox, S.; Karnawat, K.; Mydland, M.; Dumais, S.; and White, T. 2005. Evaluating implicit measures to improve web search. *ACM Transations on Information Systems* 23(2):147–168.

Freund, Y.; Iyer, R.; Schapire, R. E.; and Singer, Y. 1998. An efficient boosting algorithm for combining preferences. In *Proceedings of International Conference on Machine Learning (ICML)*.

Granka, L.; Joachims, T.; and Gay, G. 2004. Eye-tracking analysis of user behavior in www search. In *Poster Abstract, Proceedings of the Conference on Research and Development in Information Retrieval (SIGIR)*.

Granka, L. 2004. Eye tracking analysis of user behaviors in online search. Master's thesis, Cornell University.

Herbrich, R.; Graepel, T.; and Obermayer, K. 2000. Large margin rank boundaries for ordinal regression. In et al., A. S., ed., *Advances in Large Margin Classifiers*, 115–132.

Hinkelmann, K., and Kempthorne, O. 1994. *Design and Analysis of Experiments: Volume 1: Introduction to Experimental Design*. John Wiley & Sons.

Joachims, T.; Granka, L.; Pang, B.; Hembrooke, H.; and Gay, G. 2005. Accurately interpreting clickthrough data as implicit feedback. In *Annual ACM Conference on Research and Development in Information Retrieval (SIGIR)*.

Joachims, T. 2002. Optimizing search engines using clickthrough data. In *Proceedings of the ACM Conference on Knowledge Discovery and Data Mining (KDD)*.

Kelly, D., and Teevan, J. 2003. Implicit feedback for inferring user preference: A bibliography. *SIGIR Forum* 32(2).

Kemp, C., and Ramamohanarao, K. 2002. Long-term learning for web search engines. In *Proceedings of the 6th European Conference on Principles and Pratice of Knowledge Discovery in Databases (PKDD)*, 263–274.

Miller, C. S., and Remington, R. W. 2004. Modeling information navigation: Implications for information architecture. *Human-Computer Interaction* 19:225–271.

Radlinski, F., and Joachims, T. 2005. Query chains: Learning to rank from implicit feedback. In *Proceedings of the ACM Conference on Knowledge Discovery and Data Mining (KDD)*.

Teevan, J.; Dumais, S. T.; and Horvitz, E. 2005. Beyond the commons: Inversitating the value of personalizing web search. In *Workshop on New Technologies for Personalized Information Access (PIA 2005)*.

Voorhees, E. M. 2004. Overview of TREC 2004. In *Proceedings of the 13th Text REtrieval Conference*.

White, R. W.; Ruthven, I.; and Jose, J. M. 2005. A study of factors affecting the utility of implicit relevance feedback. In *Annual ACM Conference on Research and Development in Information Retrieval (SIGIR)*.

Yu, S.; Yu, K.; and Tresp, V. 2005. Collaborative ordinal regression. In *Proceedings of the NIPS 2005 Workshop on Learning to Rank*.